# Deep Subwavelength Topological Edge State in a Hyperbolic Medium

**Authors:** Lorenzo Orsini[1], Hanan Herzig Sheinfux[1], Yandong Li[2], Seojoo Lee[2], Gian Marcello Andolina[3], Orazio Scarlatella[4], Matteo Ceccanti[1], Karuppasamy Soundarapandian[1], Eli Janzen[5], James H. Edgar[5], Gennady Shvets[2] and Frank H. L. Koppens[1,6,*]

**Affiliations:**

[1]ICFO-Institut de Ciencies Fotoniques; 08860 Castelldefels (Barcelona), Spain.
[2]School of Applied and Engineering Physics, Cornell University; Ithaca, New York 14853, USA.
[3]JEIP, USR 3573 CNRS, Collège de France, PSL Research University; 11 Place Marcelin Berthelot, F-75321 Paris, France.
[4] Cavendish Laboratory, University of Cambridge, Cambridge, CB3 0HE, United Kingdom
[5]Tim Taylor Department of Chemical Engineering, Kansas State University; Durland Hall, Manhattan, KS 66506-5102, USA.
[6]ICREA-Institució Catalana de Recerca i Estudis Avançats; 08010 Barcelona, Spain.

*Corresponding author. Email: frank.koppens@icfo.eu

**Abstract:**

Topological nanophotonics presents the potential for cutting-edge photonic systems, with a core aim revolving around the emergence of topological edge states. These states are primed to propagate robustly while embracing deep subwavelength confinement that defies diffraction limits. Such attributes make them particularly appealing for nanoscale applications, where achieving these elusive states has remained challenging. We unveil the first experimental proof of deep subwavelength topological edge states by implementing periodic modulation of hyperbolic phonon polaritons within a Van der Waals heterostructure. This finding represents a significant milestone in the field of nanophotonics, and it can be directly extended to and hybridized with other Van der Waals materials in various applications. The extensive scope for material substitution facilitates broadened operational frequency ranges, streamlined integration of diverse polaritonic materials, and compatibility with electronic and excitonic systems.

**Main Text:**

Topological physics is concerned with systems whose behaviour is governed by a topological invariant, a quantity that can only attain discrete values. These invariants remain unchanged under continuous transformations of the system's parameters, rendering topological systems both robust and fundamentally interesting. While originally studied in condensed matter physics[1–3], topological behaviour is now known to be ubiquitous in wave physics and is extensively studied in photonics[4–16]. Recently, there has been a surge of interest in applying topological photonic concepts at the nanoscale[17–19]. However, the physical scales of the studied structures have thus far remained on the order of the vacuum wavelength.

Miniaturizing topological properties into the deep subwavelength regime presents a significant challenge but also offers exciting opportunities. This miniaturization requires reaching and improving the ultimate spatial limits of electromagnetic field manipulation. Additionally, it enables the exploration of how topological properties will manifest in the subwavelength (polaritonic) regime where contributions from nonlinearities[20–23], non-locality[24–26], and multimodal interactions are prominent. Furthermore, topological systems' inherent robustness and protection can be harnessed to develop more resilient deep subwavelength optical components. For instance, this includes the realization of nanocavities with a topologically fixed resonant frequency or fabrication-disorder tolerant waveguides.

Experimentally, progress has been made with plasmon polaritons in chains of metallic particles[27–29]. However, high momentum plasmon polaritons in metals always show high optical absorption, which is detrimental to the formation of polaritonic band structure. This limitation has restricted the progress done in[27–29] to wavelength-scale unit cells rather than deep subwavelength ones. An alternative platform for studying topological polaritonics is graphene, where relatively low-loss plasmon polaritons[30–32] the mid-IR show relatively low optical absorption, especially under cryogenic temperatures[33]. By electrostatically modulating graphene, it is possible to produce deep subwavelength polaritonic crystals[34]. However, the realization of topological states in such devices has proven challenging, in part due to fabrication difficulties and the need for cryogenic measurements. To address these challenges, we employed hyperbolic phonon polaritons (HPPs)[35–38], which occur in a limited frequency range in the mid-IR (the Restrahlen band) and exhibit very low absorption even at room temperature[39]. Complicated hBN polaritonic nano-devices can be relatively easily fabricated using the recently developed substrate patterning scheme, which has been applied to produce polaritonic devices such as deep subwavelength cavities[40] and simple (topologically trivial) lattices[41]. However, the multimodal nature of HPPs raises intriguing questions about whether fully developed bandgaps will form and, consequently, whether topological edge states within those gaps can be observed at all.

Here, we demonstrate for the first time a deep subwavelength topological edge state in a nanophotonic system. By employing a combination of indirect patterning techniques[34,42,43] and high-quality HPPs, we effectively mitigate the optical losses, overcoming the negative impacts of absorption and lithography-induced imperfections. Through sensitive near-field spectroscopy, we achieve the tightest confinement of the topological edge state within a geometrical volume of $0.02\ \mu m^3$, which is four orders of magnitude smaller than the illumination free-space wavelength volume of $\lambda^3 \approx 300\ \mu m^3$. Additionally, we demonstrate that a single-mode model accurately captures the expected trend for our 1D topological photonic systems. This contrasts with the complex nature of band structures found in hyperbolic systems[41]. Therefore, our findings serve as a starting point for a deeper exploration of band topology in those multimodal hyperbolic systems.

In this work, we designed 1D polaritonic lattices[44,45] utilizing mechanically exfoliated[46] isotopically pure hBN[39]. We performed scattering-type near-field optical microscopy (sSNOM)[47] and compared the experimental results with simulations to interpret the complex near-field signal. Subsequently, we computed the topological invariants of the system and crosschecked them with the observed edge states. Finally, we systematically tuned our system through a topological phase transition to study the distinctive signatures of 1D topological systems.

The general design of our 1D polaritonic lattice is shown in Fig. 1a and consists of sharply defined rectangular holes that are periodically milled through a 10 nm gold film. Then, a few tens of nanometres thick hBN flake that hosts the HPPs is placed over those holes. The dispersion relation of the HPPs changes due to the periodic modulation of the dielectric environment beneath the flake, resulting in the emergence of a polaritonic band structure. Depending on the specific design of the unit-cell, the polaritonic bands may exhibit non-trivial topology. The unit-cell, illustrated in the black-framed inset in Fig. 1a, comprises two holes separated by different spacing ($a$ and $b$). If we consider the holes in our system as cavities, the HPPs mediate the coupling between adjacent cavities by propagating through the spacing as waves. The staggered lattice spacing (a and b) naturally generates staggered hopping parameters, as in the case of a simple SSH model[44] virtue of this analogy, we refer to our 1D HPPs system as an SSH lattice.

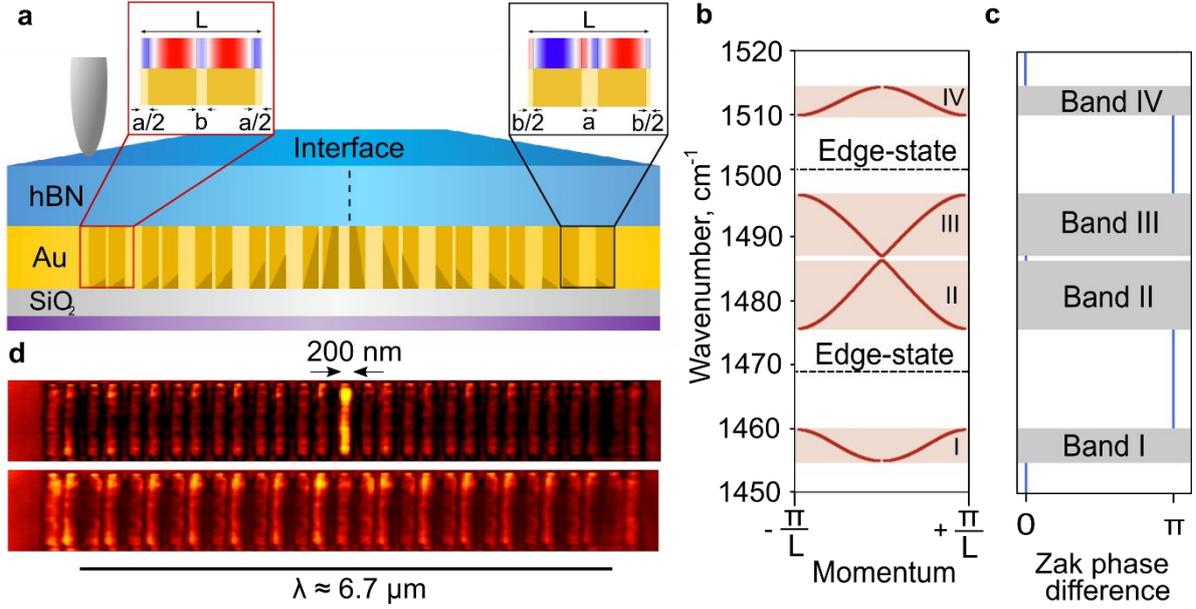

**Fig. 1 Device schematic and first experimental near-field measurements. a**, Schematic of the cross-section of the 1D polaritonic nanostructure composed of two adjacent SSH lattices. The hBN flake sits onto a periodically patterned Au film evaporated over a SiO$_2$/Si substrate. The dashed line represents the physical boundary between the two adjacent SSH lattices. In the insets, we can inspect the unit-cell of both SSH lattices and see how the inter- and intra-cell distances are swapped relative to each other. The insets also show the real part of the polaritons' electric field profile calculated at $1497\ cm^{-1}$ for the experimental conditions of Dev-01. This is calculated using the effective refractive index scheme discussed in Supplementary S2.1. The left inset shows an even field profile, whereas the other one shows an odd field profile. In the color scheme, red and blue represent positive and negative values of the electric field, respectively, while zeros are depicted in white. Above the hBN flake surface is illustrated the presence of the sSNOM tip. **b**, TMM calculated band structure of the sub-systems presented in panel **a**. The red curves are the polaritonic bands corresponding to the experimental conditions of Dev-01. The two horizontal dashed black lines represent the TMM calculated[45] edge states of the system composed of two adjacent SSH lattices. **c**, TMM calculated Zak phase difference of the sub-systems presented in panel **a**. In the gapped regions, the Zak phase difference between the left and right sub-systems can be either 0 or π. The edge states presented in panel **b** match the condition of the formation of topological edge states, for which the Zak phase difference is π. **d**, Experimental near-field spatial scan at $1495\ cm^{-1}$ (top) and at $1481\ cm^{-1}$ (bottom) illumination frequencies. The scale bar matches the wavelength at resonance and highlights the deep subwavelength nature of the bright near-field feature.

By arranging two SSH lattices side by side, such that the distances $a$ and $b$ characterizing the right lattice are swapped with respect to the left one (see both insets of Fig. 1a), a topological edge state is expected to form at the physical boundary where the two lattices meet. In fact, the theory of band topology predicts that, in chiral symmetric 1D systems, topological edge states will arise at an interface between two subsystems that have a shared energy bandgap and have different topological invariants (Zak phases). To predict the bulk band structure of our system, we employed the transfer matrix method (TMM) with periodic boundary conditions (see Supplementary section S2.1). For the prediction of the edge states, we computed the Zak phases directly by leveraging a previous theoretical study[45], which established a relationship between these phases and the lattice reflection coefficient in a generic 1D lattice. The calculation of the reflection coefficient can also be performed using the TMM. Fig. 1b depicts the simulated polaritonic band structure and illustrates the occurrence of two edge states within distinct gaps. Notably, these edge states emerge only when a non-zero difference exists between the Zak phases of the two lattices, as shown in Fig. 1c. Specifically, we simulated a 38 nm thick hBN with a unit-cell length of 542 nm and inter-/intra-cell distances, respectively, of 81 nm and 55 nm. The conditions correspond to the experimental parameters of the lattice named "SSH 13" belonging to Dev-01 (Supplementary S3.1 and S3.2).

To probe these topological edge states experimentally, we studied Dev-01 and employed sSNOM for characterization. This nano-imaging technique is well suited for probing polaritons in stationary conditions. By illuminating the near-field probe, the latter excites the HPPs within the hBN flake, which propagate through the 1D nano-structure. Consequently, the electromagnetic field associated with these HPPs accumulates and eventually reaches a stationary state. The sSNOM tip out-couples these stationary waves into the far-field, enabling us to probe features such as polaritonic bands and edge states.

Using this technique, we measured the near-field response of the sample at various illumination wavelengths. As shown in Fig. 1d, when the illumination wavelength is tuned inside the non-trivial gap around $1495\text{ cm}^{-1}$, a prominent bright feature emerges precisely at the interface between the two SSH lattices. Importantly, this feature exhibits deep subwavelength localization, confined in a single hole within the nano-structure with a lateral size of $\approx 200$ nm. As we will demonstrate throughout the paper, this feature is associated with a topological edge state, which disappears when the illumination wavelength is tuned into the dispersive bands around $1480\text{ cm}^{-1}$ (Dev-01 band structure is shown in Fig. 1b), resulting in a weaker and delocalized response originating from the entire nanostructure.

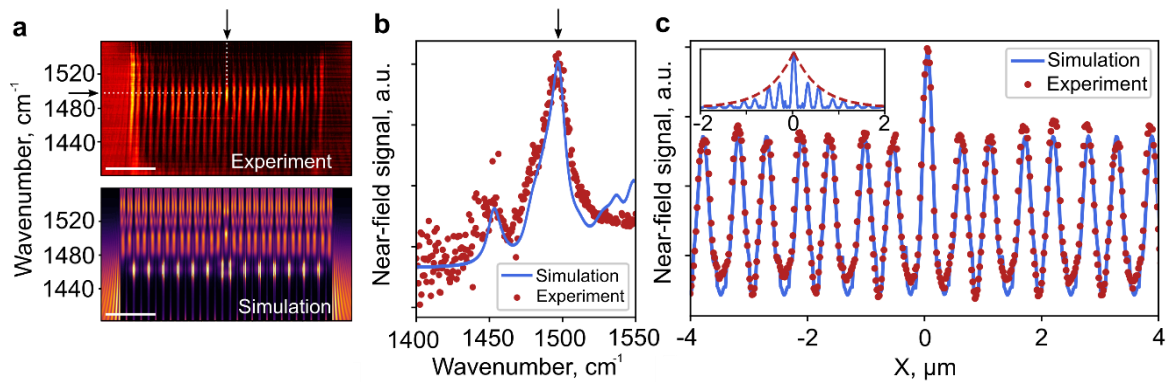

**Fig. 2 Edge state and near-field characterization. a**, Top: experimental pseudo-heterodyne near-field image of the lattice. The measurement is taken by scanning back and forth the lattice longitudinally (x-axis), and at each repetition; the illumination wavelength is incrementally changed (y-axis). The bright near-field spot, marked with the two black arrows, is located at the interface between the two adjacent SSH lattices. The scale bar corresponds to 2μm. Bottom: Simulated pseudo-heterodyne near-field image of the lattice. We can identify many of the distinctive features of the experimental near-field image plotted above, such as the localized bright feature and the delocalized band. The scale bar corresponds to 3μm. **b**, Experimental and simulated spectral response of the bright near-field feature. The resonance peak is marked with the black arrow above the plot. This is the vertical cross-section of the near-field images in panel **a** taken at the interface marked with the vertical black arrow. **c**, Experimental and simulated spatial response of the bright near-field feature. This is the horizontal cross-section of the near-field simulations presented in Supplementary Fig. S5a at resonance. Inset: Simulation of the edge state at resonance in the reduced loss case (ten times lower than the analytically calculated losses). The dashed line is an exponential decay function and highlights the localized character of the edge state. The inset's x-axis scale is μm.

To better characterize the edge state, we first performed a comprehensive characterization by simultaneously collecting the entire lattice's spectral and spatial near-field response (Supplementary

S1.2). In this process, the sSNOM scanned the lattice longitudinally along a single line while incrementally adjusting the illumination wavelength. As illustrated in the upper part of Fig. 2a, the measurement confirms the existence of the localized edge state, which is observed as a bright feature that is spatially localized at the center of the scan and spectrally very narrow. The analysis of the near-field feature spectrum depicted in Fig. 2b reveals a prominent peak at $1498 \text{ cm}^{-1}$. Additionally, we observe a narrow full-with-half-maximum of approximately $15 \text{ cm}^{-1}$, indicating the sharp frequency response of the near-field feature.

In sSNOM measurements, a bright near-field signal does not consistently signify a resonance (edge state) of the system. Supplementary section S2.2 briefly describes the algorithm employed to simulate the near-field response of our system that can be later understood in terms of physical properties calculated with standard TMM (Supplementary S3.2 and S3.3). Fig. 2a shows a notable agreement between the experimental and simulated data, which is particularly evident within the range of $1480 \text{ cm}^{-1}$ and $1520 \text{ cm}^{-1}$. In particular, Fig. 2b highlights the perfect overlap between the simulated spectral response (visible as a blue solid line) and the experimental data. As elaborated further in Supplementary section S3.3, the near-field response shown in Fig. 2a-b exhibits a strong correlation with the calculated bands and edge state of the system (Fig. 1b). Moreover, the topological origin of the edge state was previously demonstrated in Fig. 1c, which shows the calculated Zak phases under our specific experimental conditions. The measured edge state corresponds to the one that emerges between bands III and IV in Fig. 1b. An alternative view on the topological origin of the edge state arises from the consideration that the two adjacent bulk regions with opposite Zak phases exhibit a field distribution symmetry that is opposite in the two regions. This is illustrated in the insets of Fig. 1a, displaying the calculated real part of the electric field distribution for $1497 \text{ cm}^{-1}$ (the upper limit of band III). In the left unit cell, the electric field distribution is even, whereas inside the right one, it is odd. From a physical standpoint, this displacement leads to the emergence of the SSH edge mode in accordance with the predictions of band topology.

We note that in addition to the strong localized near-field response, there is also a significant signal originating from the bulk. This is particularly evident when examining the spatial profile of the near-field response (see Fig. 2c). While the response is considerably higher at the interface, it is not exponentially localized as expected from a topological edge state that emerges in the gap. Nevertheless, we show that our results are still consistent with band topology. The losses experienced by the HPPs broaden the polaritonic bands, allowing the edge state to leak into the bands. In this scenario, it is plausible that the experimental near-field response is a combination of both the edge state and the delocalized bulk states. To validate this hypothesis, we utilized the near-field TMM to simulate the edge state and observe its spatial characteristics as we approach the lossless limit. As shown in the inset of Fig. 2c, the edge state simulated with reduced losses becomes exponentially localized at the interface. This localization suggests that the edge state in this reduced-loss scenario is indeed isolated from any bulk state and is consistent with the expected behaviour of a state in a gapped system.

To provide further evidence supporting the topological origin of the observed edge state, rather than it being a different type of interface bound state, we experimentally tuned our system across a topological phase transition. For this purpose, we studied a second device (Dev-02, see Supplementary S4.1), consisting of a series of adjacent SSH lattices, which differ from Dev-01 in terms of lattice constant and parameters. As depicted in Fig. 3a, the right lattice of each set has a fixed set of parameters $(L, \bar{a}, \bar{b})$, while the left lattice maintained the same unit-cell length but exhibited different inter- and intra-cell distances ($a$) and ($b$). Similar to an SSH model, traversing a topological phase transition involves tuning the inter-cell distance to exceed the intra-cell one. It is, therefore advantageous to represent our data in relation to the "dimerization" parameter $D =$ $(a - b)/(\bar{a} - \bar{b})$. The value $D = 0$ corresponds to the critical point of the transition, while $D < 0$ in our data corresponds to a topologically non-trivial configuration.

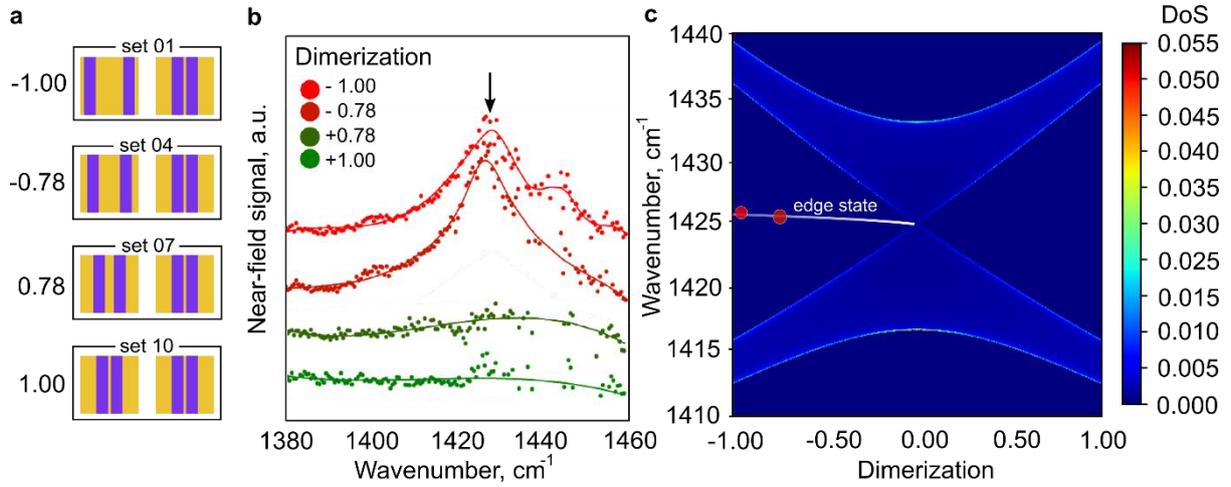

**Fig. 3 Topological phase transition characterization. a**, Illustration of the unit cells of some specific sets in Dev-02. The number to the left is the dimerization value of the specific set. Inside the boxed area, we can find the top-view representation of the unit cells of the left and right SSH lattices. The purple rectangles represent the holes milled through the gold film, whereas the golden areas are proportional to the inter- and intra-cell distances. The right SSH unit cell is the same for all sets, whereas the left SSH is slowly changing from one dimerization to the other. **b**, Experimental near-field spectra of the edge states were taken for different sets (01, 02, 09, 10). The solid lines passing through the experimental points are spline interpolations as a guide to the eye. At dimerization $-1.00$ and $-0.78$, the spectrum displays a resonance response peaked at $1426$ cm$^{-1}$, marked by a black arrow. **c,** Simulation of the density of states (DoS) of the left SSH lattice at different dimerization values. The deep blue color identifies a gapped region, whereas the lighter tone of blue corresponds to the polaritonic bands. When the system flips dimerization sign, the band gap experiences a closing followed by a reopening, as expected from the topological phase transition. The white line plotted on the left side of the map represents the theoretical topological edge state, which overlaps with the experimental observations (red circles).

Fig. 3b shows the near-field response of the interface for various sets with dimerization values ranging from -1 to 1. The measurements are conducted in a manner similar to the one depicted in Fig. 2b, albeit with a slightly modified procedure discussed in Supplementary section S4.2. We observe that the edge state abruptly disappears upon approaching positive dimerization, as expected from crossing a topological phase transition. In Fig. 3c, we plot the simulated density of states (DoS) of the left lattice, specifically focussing on the two bands involved in the transition, along with the simulated topological edge state. The frequency of the topological edge state, as predicted using TMM[45], is visible for $D < 0$ as a white line. These plots align with our experimental findings (red circles plotted in Fig. 3c), confirming that the observation of an edge state is exclusively limited to negative dimerization values. It is important to note that the DoS simulations were conducted in a lossless scenario, neglecting dissipative processes that can adversely impact the edge states. In a realistic scenario, the optical losses reduce the size of the band gap. Consequently, under real experimental conditions, the edge states are not observed near the critical point ($D = 0$), where the band gap becomes very small ($< 10$ cm$^{-1}$), strongly indicating that the band gap is already closed. Furthermore, Fig. 3c illustrates that the bands remain approximately symmetric around the energy where the gap closes, while the simulated topological edge state energy is located around the average energy of the two bands. This shows that the transition preserved an underlying chiral symmetry. This symmetry, in fact, is essential for defining topologically distinct phases in one-dimensional systems[48].

In conclusion, we experimentally demonstrated deep subwavelength topological edge states using the proposed HPPs platform. Our experimental results are consistent with previous theoretical works of band topology in one-dimensional systems and support some of the proposals in similar platforms. Moreover, our study represents a substantial advancement in the precise control of light at the nanoscale, offering an alternative platform for the realization and investigation of topological physics in nanophotonic systems. It is important to note that despite the potential challenges posed by the multimodal nature of the HPPs, we were able to observe and characterize topological phenomena.

By extending this platform to 2D systems and exploring photonic Chern insulators and the quantum valley hall physics, it becomes feasible to realize robust photonic transport down to the deep subwavelength regime. Moreover, the unique property of Van der Waals materials to stack and twist on top of each other offers an intriguing opportunity to manipulate HPPs in more complex systems and active devices. For example, the hybridization of this platform with graphene plasmon-polaritons can enable a gate-tunable topological nanophotonic system. Furthermore, the insights obtained from this research can be extrapolated to other hyperbolic materials, facilitating a broader coverage of the electromagnetic spectrum.

**Acknowledgements**

F.H.L.K. acknowledges support by the ERC TOPONANOP under grant agreement no. 726001, the Government of Spain (FIS2016-81044; Severo Ochoa CEX2019-000910-S), Fundació Cellex, Fundació Mir-Puig, and Generalitat de Catalunya (CERCA, AGAUR, SGR 1656). L.O. acknowledges support by The Secretaria d'Universitats i Recerca del Departament d'Empresa i Coneixement de la Generalitat de Catalunya, as well as the European Social Fund (L'FSE inverteix en el teu futur)—FEDER. G.M.A. acknowledge funding from the European Research Council (ERC) under the European Union's Horizon 2020 research and innovation programme (Grant agreement No. 101002955 – CONQUER). S.L. acknowledges the National Research Foundation of Korea (NRF) grant funded by the Korean government (MSIT) (Grant No. NRF-2023R1A2C1007836 and NRF-2020R1C1C1012138). J.H.E and E.J. acknowledge the support for hBN crystals growth coming from the Office of Naval Research, award number N00014-22-1-2582. O.S. acknowledges the support of the Engineering and Physical Sciences Research Council [grant number EP/W005484]. For the purpose of open access, the author has applied a Creative Commons Attribution (CC BY) licence to any Author Accepted Manuscript version arising. H.H.S. acknowledges funding from the European Union's Horizon 2020 programme under the Marie Skłodowska-Curie grant agreement ref. 456 843830. M.C. acknowledge the support of the "Presencia de la Agencia Estatal de Investigación" within the "Convocatoria de tramitación anticipada, correspondente al año 2020, de las ayudas para contractos predoctorales (Ref. PRE2020-XXXXXX) para la formación de doctores contemplada en el Subprograma Estatal de Fromación del Programa Estatal de Promoción del Talento y su Empleabilidad en I+D+i, en el marco del Plan Estatal de Investigacón Científica y Técnica de Innovación 2017-2020, cofinanciado por el Fondo Social Europeo". K.S acknowledge the support from the European Commission in the Horizon 2020 Framework Programme under Grant Agreements Nos.785219 (Core2) and 881603 (Core3) of the Graphene Flagship. G.S. and Y.L. acknowledge the support of the Office of Naval Research (Grant No. N00014-21-1-2056), the Army Research Office (Grant No. W911NF-21-1-0180), and the National Science Foundation MRSEC program (Grant No. DMR-1719875).

**Authors Contributions:**

L.O. and H.H.S. worked on sample fabrication with help from M.C. and K.S. Isotopic hBN crystals were grown by E.J. and J.H.E. Measurements and data analysis was performed by L.O. Simulations were performed by L.O. with help from Y.L. Theoretical support was provided by O.S., G.M.A., Y.L., S.L. and H.H.S. Experiments were designed by L.O., H.H.S and F.H.L.K. All authors contributed to writing the manuscript and G.S. and F.H.L.K supervised the work.

# Supplementary Information
# Deep Subwavelength Topological Edge State in a Hyperbolic Medium

## S1 Scattering type near-field optical microscopy, sSNOM

### S1.1 Setup

We measured the near-field signal of the h$^{11}$BN nano-structures using a commercially available sSNOM from Neaspec GmbH. The hyperbolic phonon-polaritons (HPPs) are probed with a mid-IR quantum cascade laser (QCL) from Daylight Solution in combination with conventional near-field probes. For preliminary and lower-resolution measurements, we used commercially available tips Pt/Ir from NanoAndMore GmbH (ARROW-NCPt-50). For high-quality measurements, we chose the NEXT-TIP S.L. (NT-TERS-E-85 Gold). The measurements were taken in a pseudo-heterodyne detection scheme using p-polarized light with powers in the range between 1 mW and 10 mW. In particular, we measured the 4$^{th}$ and 5$^{th}$ demodulation harmonic components of the reflected signal modulated with a tapping amplitude in the range between 90 nm and 120 nm.

### S1.2 Measurements

**Normalization:**

Here, we describe the method used for simultaneously collecting the entire lattice's spectral and spatial near-field response. The approach involves scanning the lattice along a single line and adjusting the QCL emission wavelength incrementally. In this context, we need to address the highly variable power spectrum issue of the QCL, which emits irregularly at different wavelengths. This variability can be observed in Fig. S1a, which shows a near-field characterization taken at the edge of the h$^{11}$BN flake belonging to Dev-01. Specifically, we can observe the presence of horizontal features due to each row of the scan being measured at a distinct emission wavelength, which corresponds to a different emission power level.

To address this issue, we divide each line by the average signal measured at the edge of the scan: a reference region with known optical properties. In this case (see Fig. S1a), the reference region is a flat gold film with no wavelength dependence in the range of interest (mid-IR). By using this normalization technique, as shown in Fig. S1b, we clearly observe the polaritonic fringes and measure the wavelength dependence of the bare h$^{11}$BN flake (the region to the left of the scan). In our experience, the bare h$^{11}$BN flake near-field response could depend on many factors, such as flake thickness, near-field probe shape, and contingent experimental conditions. Nevertheless, as shown in Fig. S1c, we characterize this dependence by observing a mild and monotonic behavior in function of the illumination wavelength.

When measuring our polaritonic lattices, it is often impractical to have a wavelength-independent normalization region in the same scan. Nevertheless, a clear signal can still be extracted using this technique. An example of a raw near-field signal coming from a lattice is shown in Fig. S1d. As expected, the scan is affected by power instability, but as shown in Fig. S1e, it is successfully normalized by using the right-hand side part of the map as a reference region. Even though this reference is not precisely wavelength-independent, we can reconstruct the normalized near-field response with an additional normalization step. The optical properties of that specific normalization region were measured before and corresponded to the data plotted in Fig. S1c. Therefore, we can reconstruct the normalized near-field map for any given scan without a wavelength-independent normalization region. For example, as shown in Fig. S1f, we reconstruct the normalized spectral response of the edge state by multiplying the signal presented in Fig. S1e by the wavelength dependence data provided in Fig. S1c.

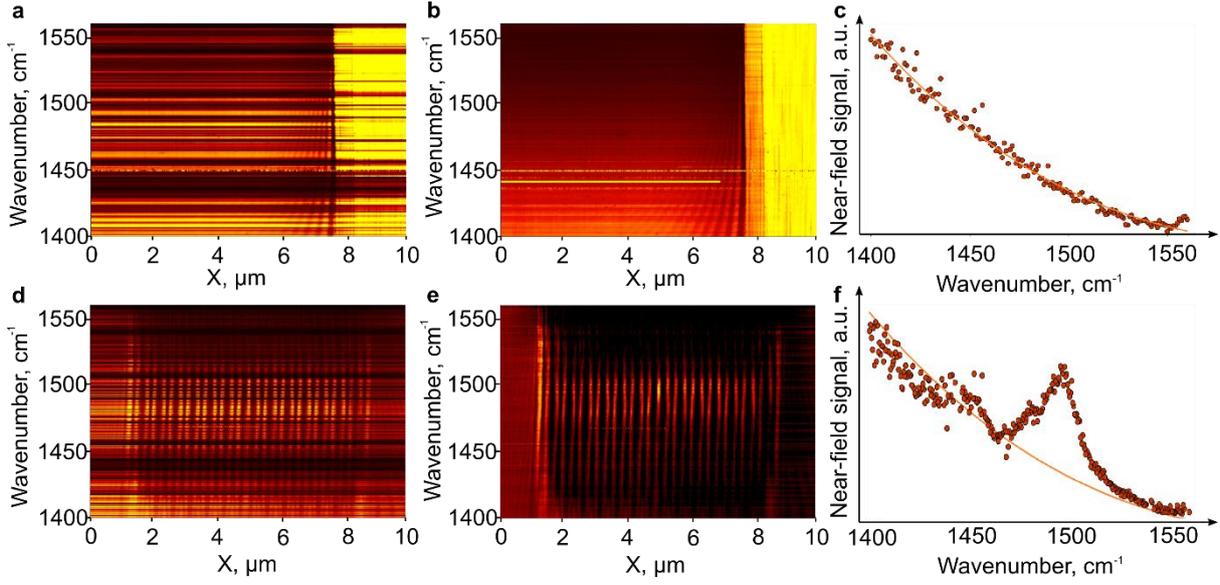

**Fig. S1 Near-field comprehensive characterization and normalization. a**, Experimental pseudo-heterodyne near-field image taken at the flake's edge marked by a white dashed line in Fig. S3b. The edge interface is located approximately at 8μm, the left-hand side of the scan is the near-field response of the h[11]BN, whereas the right one is the gold film. The horizontal dark lines correspond to low-power emission wavelengths of the QCL. **b**, Normalized near-field scan shown in **a**, the HPPs fringes are visible near the flake's edge. **c**, Vertical cross-section is taken at 1μm of panel **b**, i.e., the spectral response of the bare h[11]BN flake. The solid line is a polynomial interpolation of the experimental points. **d**, Experimental pseudo-heterodyne near-field image of "SSH 13". **e**, Normalized near-field scan is shown in **d**. **f**, Spectral response of the lattice interface plotted in E take at $\approx 5$μm multiplied by the bare near-field response of h[11]BN flake presented in panel **c**. The solid line is the bare near-field response of h[11]BN.

**Implementation:**

It is essential to mention that our measurements are done using the pseudo-heterodyne scheme without adjusting the vibration amplitude of the reference-arm mirror at any increment of the emission wavelength. In general, the vibration amplitude of this mirror is crucial to correctly measure the real and the imaginary part of the near-field signal. Precisely, the near-field reading of an sSNOM is calibrated when (see ref.[49]):

$$\Delta l \frac{4\pi}{\lambda} = \gamma = 2.63 \qquad \qquad 1$$

Here, $\Delta l$ is the mirror vibration amplitude, $\gamma$ is the modulation depth, and $\lambda$ is the illumination wavelength. In the following, we show that adjusting the mirror vibration amplitude does not critically affect the microscope calibration because, in the experiment, we tune the QCL in a relatively narrow wavelength range (7.14μm to 6.4μm).

First, we calibrated the microscope around 6.8μm, and using equation 1, we calculated our mirror vibration amplitude to be $\Delta l \approx 1.42$μm. Now, we can evaluate the discrepancy of the modulation depth $\gamma$ at different wavelengths specifically, when the illumination wavelength reaches the two extrema of the wavelength range: $\gamma_{7.14\,\mu m} = 2.50$ and $\gamma_{6.4\,\mu m} = 2.78$. Therefore, we compute the relative mismatch with respect to $\gamma = 2.63$, finding a maximum discrepancy in the modulation depth range of about 6%. Next, we must consider how the modulation depth is related to the near-field signal. In a pseudo-heterodyne measurement, the real and imaginary parts of the signal are proportional to the inverse of the Bessel function of the first kind $J_n(\gamma)$[49]. Without delving into the details of the calculation, we evaluate the real and imaginary parts of the signal at modulation depths $\gamma_{7.14\ \mu m} = 2.50$ and $\gamma_{6.4\ \mu m} = 2.78$. Provided that since our experimental observations rely solely on the near-field amplitude, we compute the module of those worst-case scenarios, and we estimate a maximum relative discrepancy in the near-field signal amplitude of around 3.9%. Such a mismatch, which maximizes at the edge of our experimental range, is definitely too small to affect the quality of our measurements negatively.

## S2 Simulation methods

### S2.1 The polaritonic propagation problem

The transfer matrix method (TMM) is widely used to solve the propagation of plane waves in one-dimensional multi-layered structures. In the case of electromagnetic waves, the method solves the one-dimensional Helmholtz equation with boundary conditions for electromagnetic waves. This method involves representing each layer as a matrix that describes the transmission and reflection of waves across that specific layer. By multiplying these matrices together, one can obtain the overall scattering matrix that describes the behavior of the wave as it passes through all the layers.

Those matrices are evaluated from the refractive index of each layer, and, in our case, we model our one-dimensional polaritonic lattice as a layered system (see Fig. S2a) by providing an effective refractive index of each section $n_{eff}$:

$$n_{eff} = \frac{\lambda_0}{\lambda_{HPP}(\lambda_0, t, \epsilon)} \qquad 2$$

Here, $\lambda_0$ is the illumination wavelength, and $\lambda_{HPP}$ is the analytically evaluated HPPs wavelength that depends on $\lambda_0$, on the h$^{11}$BN flake thickness t, and whether the flake sitting over the gold film or is suspended over the milled hole of the nanostructure (see Fig. S2a insets). With this approach, we calculate quantities such as the transmission and reflection coefficients and compute relevant properties of our polaritonic lattices, such as the Zak phase and the band structure.

### S2.2 The near-field TMM

**The near-field problem**

In the main text, we refer to an algorithm based on a variation of the TMM capable of solving the near-field response of our one-dimensional polaritonic systems. In the following, we review the central concept behind our approach, starting by breaking down the polaritonic near-field interaction of our system as a three-step process:

1. Light couples into the structure at the tip location.
2. The polaritons propagate back and forth through the structure.
3. The polaritons couple out from the structure at the tip location.

As shown in Fig. S2b, this whole process can be intuitively modeled by introducing a scattering element with an extra transmission/reflection channel in which the incident far-field light couples to the near-field polaritons and vice versa. This is standard practice in microwave electronics[50], where complex devices with multiple input/output ports are modeled as NxN scattering matrices, where N is the number of ports. In our case, we model the near-field probe as a 3x3 scattering matrix to include the extra transmission/reflection channel. That is because the polaritons can travel below the tip apex inside the hBN flake (1$^{st}$ and 2$^{nd}$ ports), and they can eventually couple with the far-field (3$^{rd}$ port).

We extend the standard TMM formulation to incorporate this scattering matrix and simulate the scanning process by computing the far-field reflection amplitude and phase of our polaritonic system at any given point. This enables us to effectively simulate the scanning process and calculate the near-field response of the polaritonic system when the tip is placed at any given location.

Without delving into the details of the formulation, the near-field probe 3x3 scattering matrix is derived by applying physical symmetries such as time-reversal symmetry, mirror symmetry, and energy conservation. In addition, the coupling between the incident light and the polaritons is modeled by taking into account its nonlinear dependence with respect to the tip-substrate distance. Furthermore, the model also accounts for the finite radius of the tip. Further details of this method are given in[51].

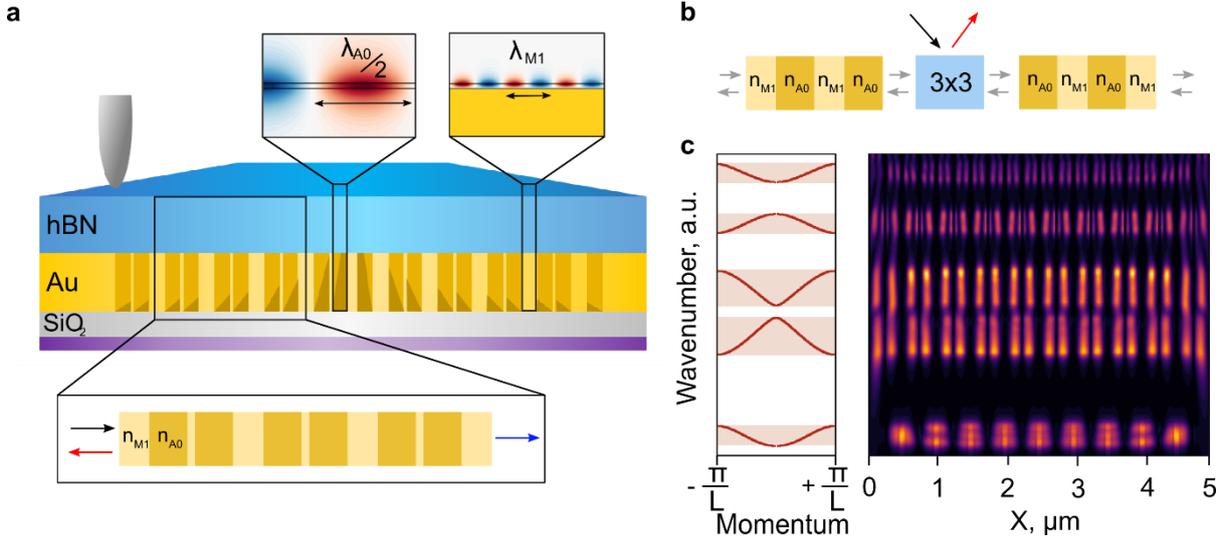

**Fig. S2 TMM polaritonic propagation and near-field observable models. a**, Schematic of the cross-section of the 1D polaritonic nanostructure composed of two adjacent SSH lattices. The two insets above the main figure show the wavelength of HPPs in different environmental conditions. To the left, there is the response of HPPs propagating through a section of $h^{11}BN$ suspended over a hole in the nano-structure. The right-hand side inset shows the same HPPs propagating when the flake is in direct contact with a gold substrate. We observe a dramatic difference in the HPPs wavelength. The inset below the main figure represents the TMM "block diagram" visualization of the enclosed section of the lattice. Blocks with different lengths and different effective refractive indexes model each milled/un-milled section. **b**, Block diagram representation of the near-field scattering process in terms of TMM. The blue block represents the near-field probe. The arrows on this block represent the incident (black) and the scattered (red) fields. **c**, Comparison between TMM simulated bands (to the left) and the "near-field TMM" response of the same polaritonic lattice (to the right). We observe a perfect match between the bands and the delocalized bright features.

**The near-field observable**

The far-field reflection amplitude and phase calculated with this TMM map directly to the scattered electric field measured at the detector in sSNOM experiments. Therefore, to evaluate the near-field observables, we need to simulate our system at many tip-substrate distances, mimicking the fact that the near-field probe undergoes subwavelength harmonic oscillations above the substrate. Afterward, we compute the near-field observable ($O_n$) by evaluating the $n^{th}$ Fourier coefficient of the simulated time-dependent scattered electric field ($E^s(t)$). This is done by considering the following integral:

$$O_n = \frac{2}{T}\int_{-T/2}^{T/2} |E^s(t)|^2 \cdot e^{i(2\pi n)\cdot \Omega t} dt \qquad 3$$

Here, $\Omega$ is the tip oscillation frequency, n is the harmonic order, and T is the tip oscillation period. Fig. 2Sc shows an example of this technique by comparing the band structure calculated with standard

TMM with the predicted near-field response of the same polaritonic lattice. We observe a precise match between the standard TMM results and the simulated near-field response.

# S3 Device 01

## S3.1 Fabrication & Nano-characterization

**(1) Metal film evaporation:**

The fabrication of Dev-01 begins with the evaporation of a 2nm Cr film, followed by the evaporation of a 10nm Au film onto a silicon nitride (SiN) membrane. The commercially available silicon nitride membranes from Norcada were chosen for their specific nominal size and thickness (100×100 μm, 10nm/20nm), which we found crucial for a good device yield. The thin metal film evaporation was done using Kurt J. Lesker Company's LAB 18 Thin Film Deposition System, with the slowest deposition rate recipes at a high vacuum to minimize the roughness of the films (> 500 ppm RMS). Specifically, the Cr layer was evaporated at 1 Å/s and the Au layer at 0.5 Å/s.

**(2) Focused ion beam lithography:**

The second step involves milling the nanostructures through the membrane using focused ion beam (FIB) lithography. We used Zeiss Orion, which can operate with He$^+$ ions and allows the milling of nanoscale features as small as a few nm. In general, He$^+$ ions tend to accumulate in pockets below the substrate and warp the milled nanostructure. Therefore, it is essential to use a membrane thinner than the penetration length of the He$^+$ ions to avoid the formation of such pockets.

**(3) Hexagonal boron nitride exfoliation:**

The third step is the mechanical exfoliation of isotopically pure h$^{11}$BN obtained through collaboration. A thin crystal of h$^{11}$BN is pre-exfoliated on thin Polydimethylsiloxane (PDMS) sheets, which are commercially available at Gelpak. In our experience, DGL-type films with retentions x0 and x4 enhance the yield. The exfoliation is completed over a silicon/silicon dioxide chip cut from a wafer available at University Wafer. The silicon dioxide thickness is 285 nm, enhancing the optical contrast between the substrate and flakes.

**(4) Hetero-structure fabrication:**

The last step is the dry transfer technique that allows the fabrication of the h$^{11}$BN/membrane heterostructure. The stamp is fabricated by covering a diamond-shaped piece of thick PDMS with a thin film of Polypropylene carbonate (PPC); everything is placed over a glass slide. First, the PPC stamp is slowly brought into contact with a chip covered with flakes exfoliated at the previous step. The chip was pre-heated at 60°C. When the PPC covers the selected h$^{11}$BN flake, the chip is cooled down to 30°C, and the stamp is lifted. In this way, we picked the flake, which was repeated with the same stamp to pick the membrane. This step is especially crucial because the membrane is not a van der Waals material, and there may be problems in fabricating the heterostructure. So, some structural cuts are made during FIB milling to facilitate the picking process. Finally, the heterostructure fabrication process is completed by dropping the h$^{11}$BN/membrane stack over a clean silicon/silicon dioxide chip. The heterostructure is released by slowly pulling apart the PPC after touching the chip pre-heated to 60°C. If there is any leftover dirt, the device can be rinsed in chloroform for 10 minutes, followed by a bath in acetone (15min to 12h), and then an isopropanol (IPA) wash. However, it is essential never to sonicate the device since the heterostructure can fly off. The resulting device is ready to be measured (see Fig. S3a).

**Design:**

As shown in Fig. S3b, the device hosts a set of five adjacent SSH lattices with nominal parameters (L, a, b) reported in Table 1. We remind that the left- and right-hand side SSH differ only because they swap the parameters (a, b) relative to the unit cell. The reason we chose those specific values leads back to preliminary simulations.

| Name | Nominal $L$, nm | Nominal $a$, nm | Nominal $b$, nm |
|---|---|---|---|
| SSH 14 | 490 | 60 | 30 |
| **SSH 13** | **481** | **54** | **27** |
| SSH 12 | 472 | 48 | 24 |
| SSH 11 | 463 | 42 | 21 |
| SSH 10 | 454 | 36 | 18 |

**Table S1. Dev 01 nominal lattice parameters.**

**The thickness of the hBN flake:**

After dropping the h$^{11}$BN/membrane heterostructure, we perform AFM over an edge of the flake to evaluate the thickness of the h$^{11}$BN that reads $32 \pm 2$ nm (See Fig. S3c). However, we can comment that this measurement only gives an approximate evaluation of the flake thickness since the scan is performed across two different substrates: h$^{11}$BN and gold. Those substrates may have different responses to the AFM probe, and usually, the thickness reading is affected by an offset of a few nanometers. Finally, from the optical image of the device presented in Fig. S3a, we can assume that the flake has a uniform thickness due to the homogeneity of its color.

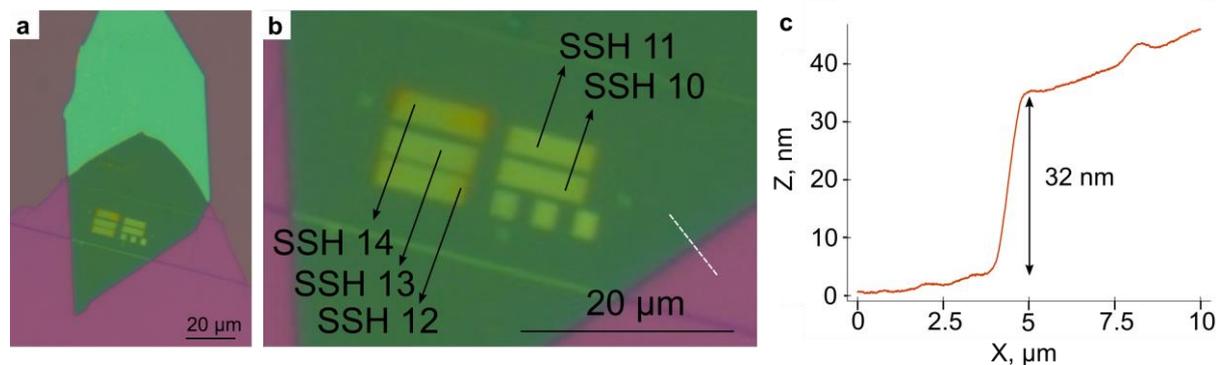

**Fig. S3 Device 01. a,** Optical picture of Dev-01. The green area is the h$^{11}$BN flake, which sits above the nano-patterned membrane (dark green) and the silicon dioxide substrate (light green). **b,** Zoom in on the nano-patterned area of panel **a**. The arrows link each pattern with its corresponding name. **c,** AFM scan taken along the white dashed line shown in panel **b** measuring the thickness of the flake.

## S3.2 Near-Field TMM of SSH 13

Here, we describe how we obtain the near-field response shown in Fig. 2 of the main text.

**Tuning the model:**

The process starts with fitting the HPPs fringes measured at the flake's edge, at the exact location as the AFM scan depicted in Fig. S3c. Fig. S4a displays those HPPs fringes, while Fig. S4b shows the plot of the fast Fourier transform of the latter. In this plot, we observe the experimental dispersion relation of the HPPs. By comparing the analytically calculated dispersion relation at different h$^{11}$BN thicknesses to the experimental one, we can find the best-fitting thickness, which is $38$ nm. The fringes are assumed to be spaced by $\lambda/2$ of the HPPs wavelength, and this assumption is valid because the damping of this type of fringes scales as $1/\sqrt[2]{x}$, as shown in Fig. S4c. Then, we benchmarked our near-field TMM by simulating this simple scenario, as shown in Fig. S4d, and observed that the simulated

HPPs fringes match in wavelength the experimental one. However, the decay length and the phase of those simulated fringes do not agree with the measured one.

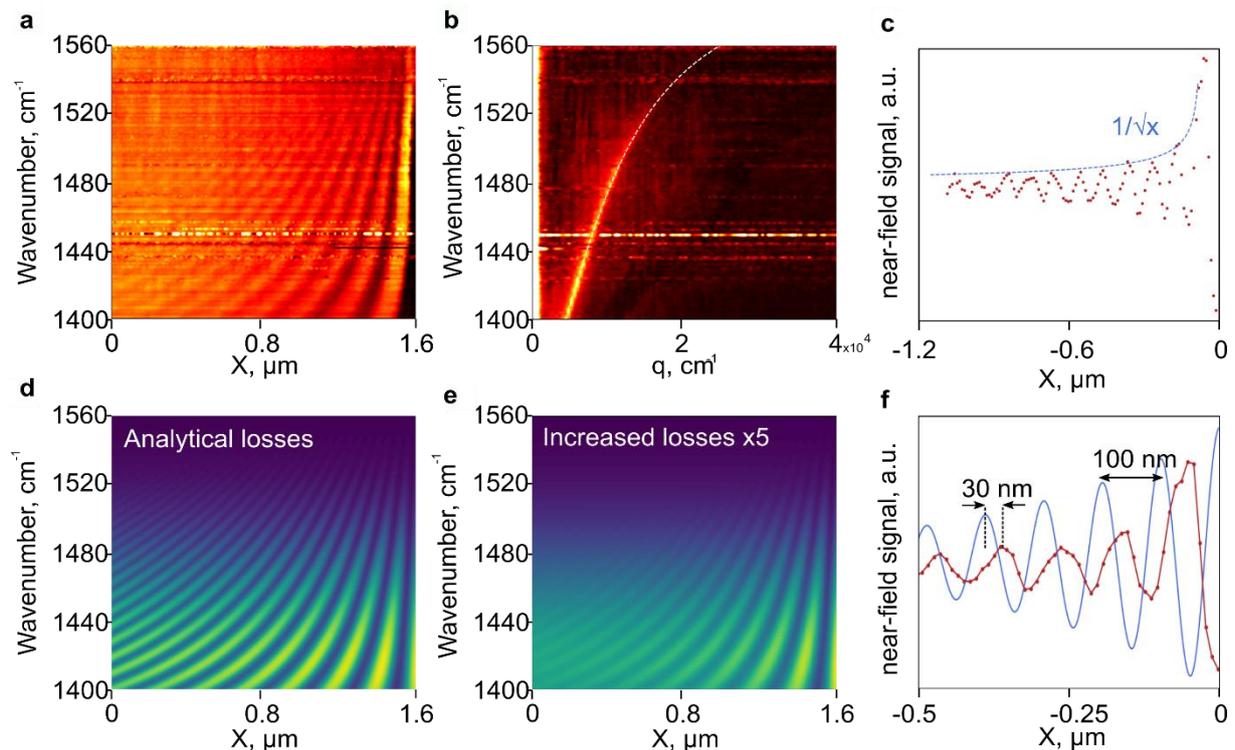

**Fig. S4 Fit of the HPPs fringes. a,** Normalized pseudo-heterodyne near-field image taken at the flake's edge (see the white dashed line in Fig. S3b). We can observe the HPPs standing waves forming from the right side of the scan. **b,** Fast Fourier transform of panel **a** performed line-wise. The bright feature that crosses the map corresponds to the experimental dispersion relation of the HPPs. The white dashed line that overlaps that feature is the analytically calculated dispersion relation for a 38 nm thick h$^{11}$BN flake. **c,** Experimental points taken from the horizontal cross-section of the scan presented in panel **a** at 1480 cm$^{-1}$. We can observe the geometrical decay rate of this standing wave, as highlighted by the blue dashed line. **d,** Near-field TMM simulation of **a**. This simulation is done using the analytically calculated losses for the HPPs. **e,** Simulation of A with increased losses. The matching with panel **a** of the HPPs standing waves improved. **f,** Comparison between the horizontal cross-sections presented in panel **a** (red dot/line) and **e** (blue solid line) taken at 1480 cm$^{-1}$. We observe how the standing waves are displaced one with respect to the other, highlighting the presence of a reflection phase that is not captured by the simulations.

We show that these discrepancies are due to the limitations of our model. Firstly, the near-field TMM simulates perfect one-dimensional systems, which does not correspond to our experimental case. The measured fringes are affected by geometrical losses caused by the tip launching the HPPs in a two-dimensional flake. Therefore, a genuine one-dimensional model cannot capture such losses. We correct our model by artificially increasing by a factor 5 the optical absorption of our simulated HPPs, which gives better results in simulating the fringes (see Fig. S4e). Finally, the phase mismatch of the fringes is shown in Fig. S4f, where a 30 nm shift between the simulated and the experimental standing waves is observed. This is because, in our model, the HPPs are described using an effective material (refer to section S2.1), which does not consider the possibility of the HPPs modes picking up a phase upon reflection.

This has profound consequences regarding the response of any structure simulated with our model. Specifically, the formation of stationary waves relies on constructive interference conditions determined by the wavelength and reflection phase at each interface of the system. As such, introducing an additional reflection phase can alter the frequency at which a particular standing wave would occur. This effect has already been reported in the context of graphene plasmon-polaritons[52], and, according to the conclusion of that research, this result can, in principle, be extended to other two-dimensional systems supporting polaritons. In our case, the HPPs create standing waves within the nano-structure that can be modeled as a multi-layer Fabry-Pérot. Therefore, we can examine the effect of introducing a reflection phase by considering a simpler Fabry-Pérot resonator composed of a cavity enclosed by two mirrors. In our HPPs system, the resonance condition for such Fabry-Pérot resonator is expressed as:

$$2n\pi = 2 \cdot \left(\frac{2\pi}{\lambda_{HPP}}l + \phi\right) \quad \quad 4$$

Here, n is the mode order, $\lambda_{HPP}$ is the HPP wavelength, l is the length of the Fabry-Pérot, and $\phi$ is the phase picked up upon reflection. Specifically, the resonant condition at $\phi = 0$ can be written as $l_{\phi=0} = n\lambda_{HPP}/2$. This notation helps to rewrite the resonant length of the Fabry-Pérot as follows:

$$l = l_{\phi=0} - \frac{\phi}{2\pi}\lambda_{HPP} \quad \quad 5$$

We can see how the resonant length l at a given wavelength $\lambda_{HPP}$ is modified with respect to $l_0$ by the presence of the reflection phase. We can understand this effect as if the Fabry-Pérot resonator has a different effective length compared with the scenario $\phi = 0$. To have another perspective on the same problem, we can write equation 6 such that the resonant wavelength $\lambda_{HPP}$ is the incognita we want to calculate:

$$\lambda_{HPP} = \frac{2\pi l}{\pi n - \phi} \quad \quad 7$$

Then, we can write the resonant condition at $\phi = 0$ as $\lambda_{HPP,\phi=0} = 2l/n$, and for a general case:

$$\lambda_{HPP} = \lambda_{HPP,\phi=0}\left(1 - \frac{\phi}{n\pi}\right)^{-1} \quad \quad 8$$

We can see how the resonant wavelength $\lambda_{HPP}$ at a given length l of the Fabry-Pérot is modified with respect to $\lambda_{HPP,\phi=0}$ by the presence of the reflection phase. We can understand this effect as if the Fabry-Pérot resonator has a different resonant wavelength than scenario $\phi = 0$. With that in mind, since our model is not able to capture this additional reflection phase, we can implement an approximate correction in two equivalent ways:

1. We can simulate our HPPs structure with effective lengths instead of the physical milled lengths and quantitatively fit the experimental response: –Effective– approach
2. We can simulate our HPPs structure with the physical milled lengths and qualitatively fit the shape of the experimental response, expecting a mismatch in the illumination wavelength: –Physical– approach

**Simulating the near-field response of SSH 13:**

With that in mind, we simulate the lattice "SSH 13" response with both approaches. As illustrated in Fig. S5a, we observed that the essential features, such as the edge state and polaritonic bands, are present in both approaches. The equivalence between the two approaches is further highlighted in Fig. S5b-c, where we display the TMM-simulated band structures for both the –Effective– and –Physical– approaches, respectively. The band structures are identical, with the only difference being a frequency shift ($30 cm^{-1}$) relative to each other. Nonetheless, the key bands highlighted in the main paper (I, II, III, and IV) are approximately equivalent. The actual lattice parameters used in the simulations are provided in detail in the supplementary Table S2.

| SSH 13 | $L$, nm | $a$, nm | $b$, nm | K shift, $cm^{-1}$ |
|---|---|---|---|---|
| Nominal | 481 | 54 | 27 | |
| Fit effective lengths | 848 | 110 | 72 | 0 |
| Fit physical lengths | 542 | 81 | 55 | -30 |

**Table S2. Comparison between fitting and nominal parameters.**

To conclude, we used the –Effective– approach to simulate the spectral response presented in Fig. 2b in the main text since this approach does not produce any frequency shift. On the other hand, we used the –Physical– approach to simulate the spatial near-field response since it matches the spatial parameters of the lattice, and the simulation results can be directly compared with the spatial experimental profile of the edge state.

### S3.3 Comparison between Near-Field TMM and Standard TMM

In the main text, we refer to the complex near-field response being mapped to standard TMM. As demonstrated in Fig. S2c and Fig. S5a-b-c, we have established that the delocalized near-field response corresponds to the polaritonic bands. The following briefly describes how the mapping holds for the edge state.

In Fig. S5d, the left-hand side of the panel displays the near-field response simulation using the analytically evaluated losses. We observe the presence of three edge-states, marked as "a", "b", and "c." The right-hand side of the panel exhibits the TMM-calculated admittance of the interface, and each peak corresponds to an edge state[45]. These peaks match the near-field features, indicating a significant correlation between them and the calculated edge states.

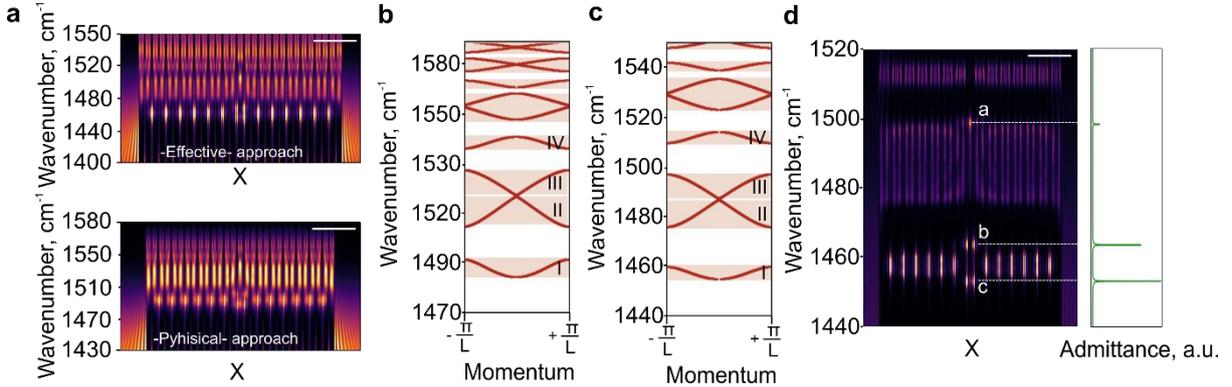

**Fig. S5 Simulations of SSH 13. a,** Near-Field TMM simulation of "SSH 13" using effective (top) and physical (bottom) lattice parameters. The main features, visible as the bright edge state and the polaritonic bands, are observable in both simulations. However, we observe a significant shift of 30 $cm^{-1}$ between the two. The scale bar in the top image of the panel is 3μm, and the scale bar in the bottom image is 2μm. **b,** Band-structure simulation of "SSH 13" using physical parameters (panel **a** bottom). **c,** Band-structure simulation of "SSH 13" using effective parameters (panel **a** top). When comparing this plot with the one in panel **b**, we observe that the relevant bands marked with I, II, III, and IV match in shape and width, and only a rigid shift in wavenumber displaces them. **d,** To the left, Simulation of "SSH 13" using the effective approach at reduced losses. This allows us to better isolate the edge states from the dispersive bulk to map the correlation of those features to the one calculated with standard TMM. The latter is shown on the right-hand side of the panel: the x-axis is the calculated admittance of the interface between the two adjacent lattices, and each peak indicates the emergence of an edge state and perfectly correlates to the presence of the near-field features. The scale bar is 2μm.

# S4 Device 02

## S4.1 Fabrication & Nano-characterization

In general, the fabrication of Dev-02 follows the same steps as Dev-01, except that the initial metal film evaporation and the FIB milling are done over a silicon/silicon dioxide chip from University Wafer.

**(1) Metal film evaporation:**

This step follows almost exactly the procedure described in Dev-01. Here, a silicon/silicon dioxide chip replaces the SiN membrane.

**(2) Focused ion beam lithography:**

This step uses a different FIB (Zeiss Auriga), which can operate with $Ga^+$ ions. In general, $Ga^+$ ions mill at a higher rate with respect to the $He^+$ ions used to fabricate Dev-01, leading to the possibility of milling larger nanostructures. However, the minimum feature size is around one order of magnitude larger than the one milled with $He^+$ ions. This change is needed since this device is designed to host a topological phase transition and requires more nanostructures.

**(3) Hexagonal boron nitride exfoliation:**

The third step is following the exact procedure described in Dev-01.

**(4) Hetero-structure fabrication:**

The last step is a simplified version of the one described in Dev-01. Specifically, we must pick the target $h^{11}BN$ flake and drop it over the milled nanostructure.

**Design:**

The device is shown in Fig. S6a, and as discussed in the main text, it comprises a set of ten lattices with AFM-measured parameters $(L, a, b)$ reported in Table S3. Those measurements are extracted from a comprehensive AFM scan shown in the inset of Fig. S6a.

| Set | $L$, nm | $a$, nm | $b$, nm | $\bar{a}$, nm | $\bar{b}$, nm | Dimerization |
|-----|---------|---------|---------|---------------|---------------|--------------|
| 01  | 740 ± 10 | 140 ± 15  | 104 ± 15  | 104 ± 15  | 140 ± 15  | -1    |
| 02  | 740 ± 10 | 136* ± 15 | 108* ± 15 | 104* ± 15 | 140* ± 15 | -0.78 |
| 03  | 740 ± 10 | 132* ± 15 | 112* ± 15 | 104* ± 15 | 140* ± 15 | -0.56 |
| 04  | 740 ± 10 | 128* ± 15 | 116* ± 15 | 104* ± 15 | 140* ± 15 | -0.34 |
| 05  | 740 ± 10 | 124* ± 15 | 120* ± 15 | 104* ± 15 | 140* ± 15 | -0.12 |
| 06  | 740 ± 10 | 120* ± 15 | 124* ± 15 | 104* ± 15 | 140* ± 15 | 0.12  |
| 07  | 740 ± 10 | 116* ± 15 | 128* ± 15 | 104* ± 15 | 140* ± 15 | 0.34  |
| 08  | 740 ± 10 | 112* ± 15 | 132* ± 15 | 104* ± 15 | 140* ± 15 | 0.56  |
| 09  | 740 ± 10 | 108* ± 15 | 136* ± 15 | 104* ± 15 | 140* ± 15 | 0.78  |
| 10  | 740 ± 10 | 104* ± 15 | 140* ± 15 | 104* ± 15 | 140* ± 15 | 1     |

**Table S3. Dev-02 lattice parameters.** The values marked with "*" are calculated considering that, by design, the sets span from dimerization -1 to +1. This constrains the values of $(L, a, b)$ for all the other sets and allows their estimation using FIB parameters since this lithography technique is more accurate than the AFM.

**Characterization:**

We characterized Dev-02 using AFM and sSNOM to measure the flake thickness. The procedure used is the same as applied to Dev-01, and the results are summarized in Fig. S6. In panel A, we observe that the flake is very homogeneous from the optical picture. In addition, the AFM scan taken at the edge of the flake reads $28 \pm 2$ nm (see Fig. S6b), which is in excellent agreement with the best fit (27 nm) of HPPs fringes measured at the same location (see Fig. S6 c-d-e).

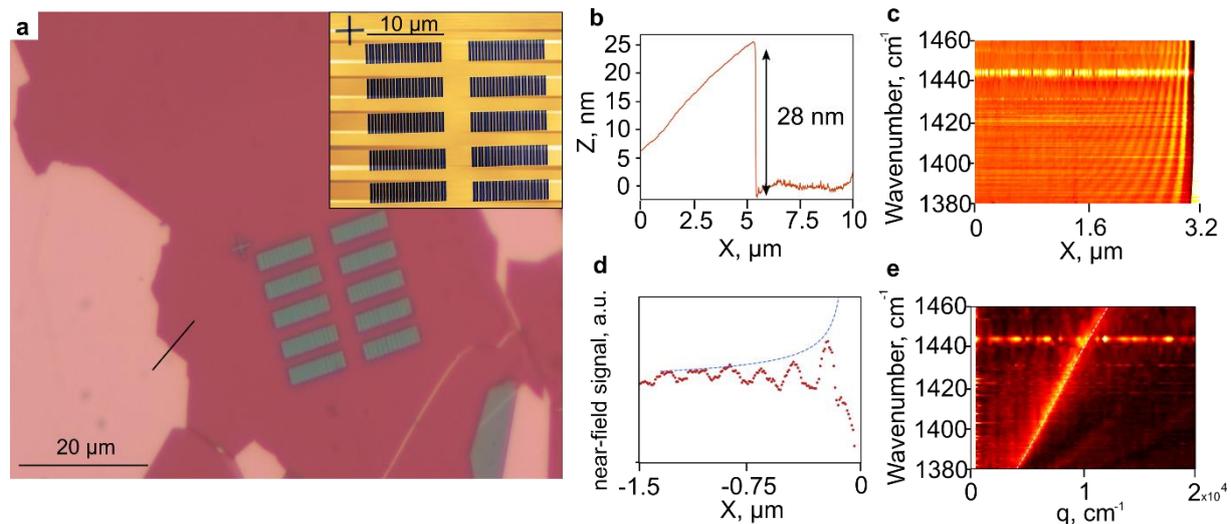

**Fig. S6 Characterization of device 02. a,** Optical image of the flake (purple) sitting over the gold substrate (pink). The flake is homogenous in thickness since the color is uniform across the image. At the center of the image, we observe the presence of the ten sets. The inset is an AFM scan of the whole set before placing the flake over the nano-patterns. **b,** AFM scan taken along the solid black line in panel **a**, measuring the thickness of the flake. **c,** Normalized pseudo-heterodyne near-field image taken at the flake's edge (see black solid line in panel **a**). **d,** The experimental points are the horizontal cross-section of the scan presented in panel **c** taken at $1385\ cm^{-1}$. This corresponds to the standing HPPs characterized by a geometrical decay rate, as highlighted by the blue dashed line that outlines the decay of the standing wave. **e,** Fast Fourier transform of panel **c** performed line-wise. The white dashed line is the analytically calculated dispersion relation for a 27 nm thick h$^{11}$BN flake.

S4.2 Near-field interface measurements

In the main text, we discussed how measuring the edge state is influenced by its interaction with the bulk states, which was also observed in the lattices in Dev-02. This interaction causes the edge state's spectrum to broaden and merge with the bulk states. Consequently, we get a mixed response when we measure the spectrum at the interface between adjacent SSH lattices (as shown in Fig. 2b of the main text).

Fig. S7a illustrates the near-field spectral response at the interface for two sets of lattices, set-01 and set-10. Surprisingly, even though only set-01 is intentionally designed to have a topological edge state, set-10 also exhibits a peaked response at the interface. However, this peaked response is due to the bulk response, not the edge state. The reason is that set-10 doesn't have an interface designed to host any edge state, so the response is solely from the bulk. Because the topological lattice spectrum contains both bulk and edge-state responses, it becomes problematic to directly compare the spectral responses at the interface between different sets. In contrast, the trivial lattice spectrum contains only the bulk response at the interface.

To isolate the edge-state spectral response, we perform a subtraction operation. We subtract the near-field response taken a few periods away from the interface (where the bulk states dominate, Fig. S7b) from the data measured at the interface (Fig. S7a). This subtraction eliminates the bulk response, giving us a clearer view of any edge state that may form at the interface. The result of this operation is shown in Fig. S7c. In this figure, it is evident that the peaked behavior, which indicates the presence of the edge state, is no longer present for set-10. On the other hand, for set-01, the peaked response is still very clear, demonstrating the effectiveness of this procedure in highlighting the presence of an edge state at the interface.

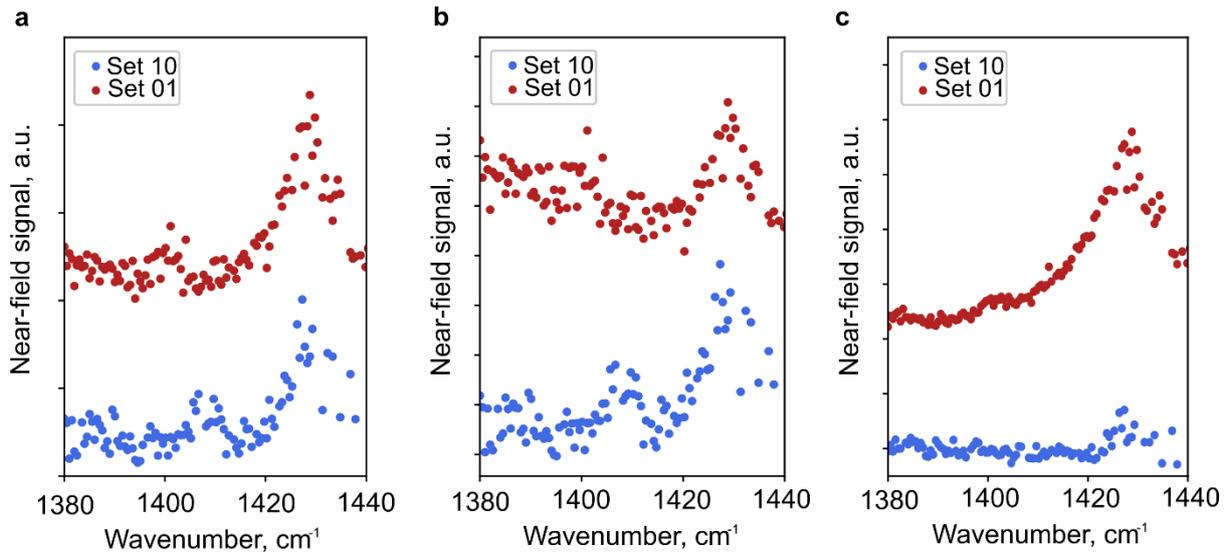

**Fig. S7 Edge-state. a,** Experimental spectral response of the lattices interface for the topological non-trivial set-01 (red) and the topological trivial set-10 (blue). **b,** Experimental spectral response of the bulk of the lattice for the topological non-trivial set-01 (red) and the topological trivial set-10 (blue). **c,** Difference between the experimental points shown in panels **a** and **b**, respectively, for set-01 and set-10. We observe how the peaked response is preserved only for the topologically non-trivial set-01.